# Theoretical investigation of two-dimensional phosphorus carbides as promising anode materials for lithium-ion batteries

*Ke Fan[a], Yiran Ying[a], Xin Luo\*[b], and Haitao Huang\*[a]*

[a]Department of Applied Physics, The Hong Kong Polytechnic University, Hung Hom, Kowloon, Hong Kong, P.R. China

[b]School of Physics, Sun Yat-sen University, Guangzhou, Guangdong Province, P.R. China, 510275

AUTHOR INFORMATION

**Corresponding Author**

*E-mail: aphhuang@polyu.edu.hk (H.H.); luox77@mail.sysu.edu.cn (X.L.)

ABSTRACT.

Employing two-dimensional (2D) materials as anodes for lithium-ion batteries (LIBs) is believed to be an effective approach to meet the growing demands of high-capacity next-generation LIBs. In this work, the first-principles density functional theory (DFT) calculations are employed to evaluate the potential application of two-dimensional phosphorus carbide (2D $PC_x$, x=2, 5, and 6) monolayers as anode materials for lithium-ion batteries. The 2D $PC_x$ systems are predicted to show outstanding structural stability and electronic properties. From the nudge elastic band calculations, the Li atoms show extreme high diffusivities on the $PC_x$ monolayer with low energy barriers of 0.18 eV for $PC_2$, 0.47 eV for $PC_5$, and 0.44 eV for $PC_6$. We further demonstrate that the theoretical specific capacity of monolayer $PC_5$ and $PC_6$ can reach up to 1251.7 and 1235.9 mAh g$^{-1}$, respectively, several times that of graphite anode used in commercial LIBs. These results suggest that both $PC_5$ and $PC_6$ monolayer are promising anode materials for LIBs. Our work opens a new avenue to explore novel 2D materials in energy applications, where phosphorus carbides could be used as high-performance anode in LIBs.

## 1. Introduction

In recent years, battery systems have made a big contribution to the rapid development of the electronics market.[1, 2] Among all types of batteries, lithium-ion batteries (LIBs), as one of the greatest successes of clean energy storage technologies, have attracted widespread attention.[3-6] Conventional anode materials in LIBs have a bulk structure with low capacity and high diffusion barrier which hinder their further applications.[7] Since the successful isolation of graphene from graphite,[8-10] two-dimensional (2D) materials, such as hexagonal boron nitride,[11, 12] phosphorene,[13-15] transition metal dichalcogenides,[16] and MXenes[17-19] have been widely studied and identified as potential anode materials with enhanced electronic properties for LIBs. Despite these efforts, designing 2D anodes with higher capacity and lower diffusion barrier which can be commercialized is still a challenging job for researchers.

It is well-known that graphene and phosphorene are two important 2D materials that have potential applications in LIBs. Graphene shows excellent conductivity and stability, but six carbon atoms can only adsorb one Li ion to form a $LiC_6$ intercalation.[20] As a comparison, one Li atom can be intercalated with two phosphorus atoms in phosphorene, while the disadvantage of phosphorene is the limited stability in air or humid environment.[21, 22] Since both elemental carbon and phosphorus are reported to form stable 2D monolayers and have similar structures including three-fold coordinated atoms and a hexagonal network,[23-25] it is reasonable to believe that compound phosphorus carbides can be stable as a monolayer and display properties that might even be superior to both constituents. Until now, several phosphorus carbide monolayers have been theoretically predicted to be stable, among which 2D PC has been fabricated successfully through doping C atoms into phosphorus ones.[25-31]. Tan et al. have synthesized metallic black phosphorus carbide (β-PC) which has been verified to be stable at the DFT level,[26] and Zhang et al. has proposed to use the γ-PC in the LIBs.[28] However, the studies of 2D phosphorus carbide as a promising anode material for LIBs are still scarce due to limited 2D candidates and possible structural instability. Recently, Yu et al. have performed extensive structural search for $PC_x$ (x=2, 3, 5, and 6) monolayers, and they found that $PC_2$ (metal), $PC_3$ (semiconductor with a band gap of 2.15 eV), $PC_5$ (metal) and $PC_6$ (semiconductor with a band gap of 0.84 eV) are thermodynamically stable.[30] The predicted $PC_x$ structures except for wide-band-gap semiconductor $PC_3$, may have potential applications in anodes for LIBs, but theoretical or experimental evidence is lacking.

Inspired by the great potential of PC in LIBs discussed above, in this work, we systematically investigate the stability, electronic and Li storage properties of $PC_2$, $PC_5$ and $PC_6$ monolayers by means of density functional theory (DFT) calculations. Our calculations reveal that $PC_x$ (x=2, 5, and 6) systems show not only excellent thermal and dynamic stability but also high electron and ionic conductivities. The relatively high capacity of $PC_5$ (1251.7 mAh g$^{-1}$) and $PC_6$ (1235.9 mAh g$^{-1}$) suggesting that both candidates could be promising anode materials for LIBs.

## 2. Computational methods

Our calculations were performed in the framework of DFT calculations by the Vienna *ab initio* simulation package (VASP).[32] The generalized gradient approximation (GGA) with the Perdew-Burke-Ernzerhof (PBE) flavor[33] was chosen as the exchange-correlation functional. Van der Waals interaction was considered by using the semiempirical DFT-D2[34] approach. For the calculation of band structures and density of states (DOS), a more accurate HSE06 hybrid functional[35] was applied. The plane-wave cut-off energy of 500 eV was employed. To avoid the interlayer interaction, a vacuum layer of 15Å was added to the slabs. The Brillouin zone was sampled using a Monkhorst-Pack k-point mesh scheme, and the meshes of Γ-centered 13×13×1 and 6×6×1 were used for the unit cell and the 2×2×1 supercell, respectively. The convergence criteria for energy and force were set to be $10^{-5}$ eV and 0.01 eV Å$^{-1}$, respectively. Charge differential analysis combined with the Bader charge method[36] was used to quantitatively estimate the charge redistribution and transfer. The climbing image nudged elastic band (CI-NEB) method[37] was applied to calculate the potential energy diffusion pathway and the minimum diffusion energy barrier of lithium ions on the $PC_x$ monolayer.

Phonon dispersion spectra was calculated by Phonopy code[38]. *Ab initio* molecular dynamics (AIMD)[39] simulations with Nosé-Hoover thermostat and NVT ensemble were performed at 300 K with the time step of 2 fs.

The adsorption energy of Li atoms on the $PC_x$ monolayer can be calculated with the following equation[40]:

$$E_{ad} = (E_{PC_xLi_n} - E_{PC_x} - nE_{Li})/n \qquad (1)$$

Here, $E_{PC_xLi_n}$ and $E_{PC_x}$ are the total energies of $PC_x$ with and without Li atom adsorption, respectively. $E_{Li}$ represents the energy per Li atom in body-centered cubic (bcc) structure, and $n$ corresponds to the number of Li atoms in the adsorption configurations.

The differential charge density is obtained as the difference between the valence charge density before and after the bonding[19]:

$$\Delta\rho = \rho_{PC_xLi_n} - \rho_{Li} - \rho_{PC_x} \qquad (2)$$

where $\rho_{PC_xLi_n}$, $\rho_{Li}$, and $\rho_{PC_x}$ represent the charge density distributions of Li-intercalated $PC_x$ systems, Li atom, and bare $PC_x$ monolayer, respectively.

The charge/discharge processes can be described by the following half-cell reactions:[31]

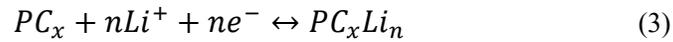

$$PC_x + nLi^+ + ne^- \leftrightarrow PC_xLi_n \qquad (3)$$

The open circuit voltage (OCV) at different coverage on the surfaces of the $PC_x$ is calculated as[41-43]:

$$OCV \approx (E_{PC_x} + nE_{Li} - E_{PC_xLi_n})/n \qquad (4)$$

The maximum capacity ($C_M$) can be obtained by the following equation[42, 44]:

$$C_M \approx zy_{max}F/M_{PC_x} \qquad (5)$$

where z, $y_{max}$, F, and $M_{PC_x}$ are the valence number, maximum adatom content (corresponding to chemical formula $PC_xLi_y$), Faraday constant (26.8 Ahmol$^{-1}$)[45], and relative molecular mass of $PC_x$, respectively. In the calculation of adsorption process, all the atoms were fully relaxed.

## 3. Results and discussion

### 3.1 Structures and stability

All three DFT-optimized structures of $PC_x$ (shown in Fig. 1 with all optimized lattice constants shown in Table S1) are quasi-planar, with P atoms located away from the basal plane. $PC_2$ with the space group of P-1 is constructed with 5-8-5 rings, similar to the case of popgraphene.[46] The 5-ring consists one P atom and four C atoms, while the 8-ring is consisted of four P and four C atoms. Each atom (P and C) is coordinated with two C atoms and one P atom. The other two structures are stabilized into two types of hexagonal structure including $C_6$, $P_2C_4$ rings for $PC_5$ and $C_6$, $PC_5$ rings for $PC_6$ (space group P-3). For $PC_5$, two types of $P_2C_4$ rings (ortho-P and para-P) are aligned in $a$ direction and $C_6$ rings fill up the plane in a zigzag configuration, and each P atom is coordinated with one P atom and two C atoms, similar to the situation of $PC_2$. For $PC_6$, on the other hand, each $C_6$ ring is surrounded by six $PC_5$ rings, and each P is coordinated with three C atoms. The quasi-planar nature of $PC_x$ may provide abundant adsorption sites for practical battery applications.

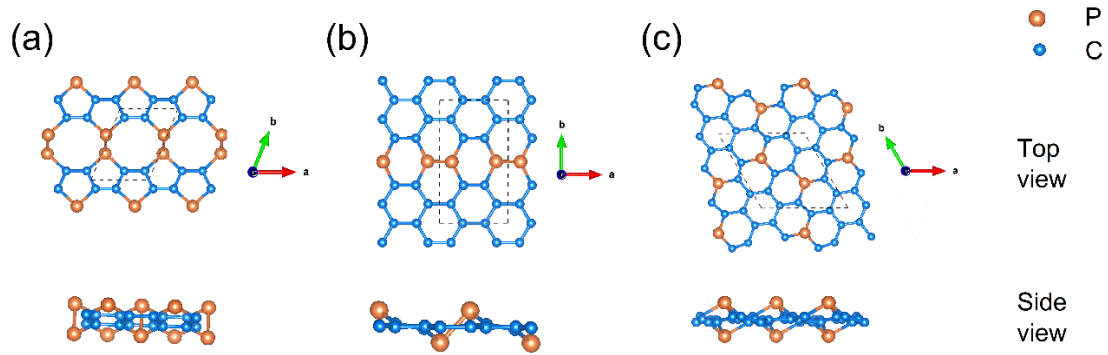

**Fig. 1** Top and side views of 2D (a) $PC_2$, (b) $PC_5$, and (c) $PC_6$. The unit cells are indicated by black dashed lines.

The structural stability is a very important factor for practical application of 2D materials. Here, we calculate the phonon dispersion to check the phase stability. Taking $PC_5$ as an example, in Fig. 2a, the absence of imaginary phonon modes in the first Brillouin zone indicates that $PC_5$ monolayer is dynamically stable. Furthermore, AIMD simulations are used to examine its thermal stability with 3×3 supercell of $PC_5$ at 300 K for 10 ps. No obvious structural reconstructions can be noticed as the total energy of the system oscillate around the equilibrium values at the room temperature (Fig. 2b). Similarly, Fig. S1 shows dynamical and thermal stability of $PC_2$ and $PC_6$.

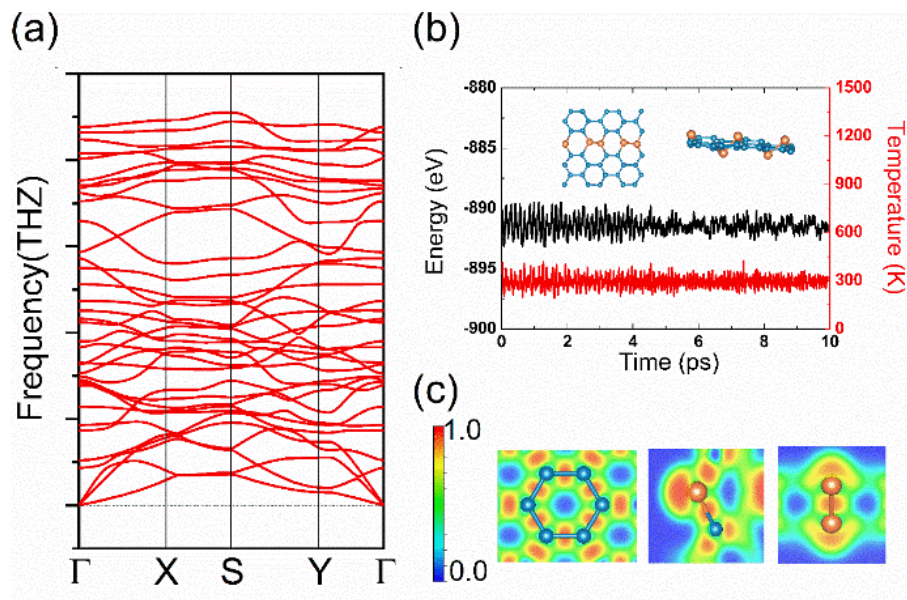

**Fig. 2** (a) Phonon dispersion spectra of 2D $PC_5$. (b) Total energy and temperature as a function of time for $PC_5$ during the AIMD simulation (inset: the structure of $PC_5$ after 10 ps AIMD simulation). (c) Electron localization function (ELF) maps of $PC_5$ along planes containing specified bonding atoms. P and C atoms are represented in orange and blue, respectively.

The calculated electron localization function (ELF)[47, 48] clearly describes the bonding behavior. Generally, ELF > 0.5 corresponds to a covalent bond or core electrons, whereas the ionic bond is represented by a smaller ELF value (<0.5). An ELF value of 0.5 is the metallic bond.[49, 50] For $PC_5$, as illustrated in Fig. 2c, the bonds between C-C, P-P and P-C atoms are covalent in nature. Each C atom bonds with the three nearest neighbors in a planar configuration implying that the C atoms are sp$^2$ hybridized. On the other hand, lone pair electron is shown near P atoms, together with the buckled configuration of $PC_5$, showing a sp³ hybridization of P atoms. This bonding configuration satisfying the chemical octet rule on both C and P sites enhances its structure stability. In $PC_2$ and $PC_6$ system (Fig. S1), C-C, P-P and C-P bonds also show covalent bond character, with sp² hybridization of C atoms and sp³ hybridization of P atoms. The detailed information about ELF plane is shown in Fig. S2.

**3.2 Electronic properties**

The electronic structure of anode material strongly correlates with the battery cyclability and rate performance. Since PBE functional usually underestimates the exact band gap values, here, we use more sophisticated hybrid functional HSE06 to perform the electronic structure calculations. Fig. 3 shows the computed density of states (DOS) and electronic band structures. Both $PC_2$ and $PC_5$ monolayers show metallic character with one Dirac cone along the Γ−Y direction (Fig. 3a and 3b) which reveal its intrinsic metallicity and high density of carriers, indicating good electronic conductivity. The major contribution to the DOS of the Dirac bands comes from the C atoms with small contribution from the P atoms. The $PC_6$ monolayer is a direct-gap semiconductor with a band gap of 0.84 eV at the M point (Fig. 3c). Despite its semiconducting nature, $PC_6$ shows high carrier mobility above $10^5$ cm² V⁻¹ s⁻¹.[30] These features render $PC_x$ as promising candidates for LIBs anodes.

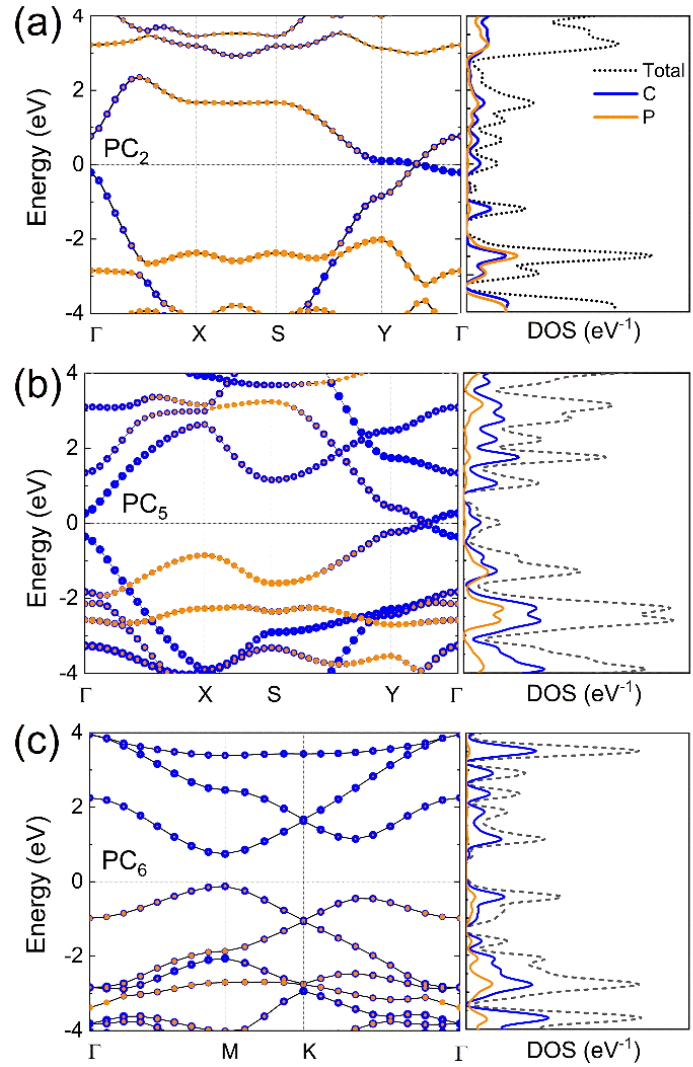

**Fig. 3** Electronic band structure (left) and PDOS (right) for 2D (a) $PC_2$, (b) $PC_5$ and (c) $PC_6$ calculated at the HSE06 level. Partial band structure contributed by P and C atoms are represented by orange and blue dots, respectively, and the sizes of the dots are proportional to the weight of the contribution. Fermi levels are set to be zero.

### 3.3 Investigation of 2D $PC_x$ as anode materials for LIBs

Aside from the stability and electronic structure, the diffusion energy barrier, specific capacity, and open-circuit voltage (OCV) are also the key performance parameters of an anode material for LIBs. To systematically study the adsorption properties, a 2×2×1 supercell is used to examine the adsorption sites for an isolated Li atom. Several possible high-symmetry adsorption sites are considered (Fig. S3). In $PC_2$ system, after full structural relaxation, only three inequivalent adsorption sites A1, A2 and A3 are identified. For Li on A1 site, the adsorption energy is -0.94 eV, while for A2 and A3 site, the values are -0.44 eV and -0.87 eV, indicating that A1 is the most favorable adsorption site (Fig. S3a). The adsorption energies are -0.84 and -0.83 eV (Fig. S3b and S3c) for the most favorable A1 sites of $PC_5$ and $PC_6$, respectively. The relatively large adsorption energies of Li atom on $PC_x$ systems ensure the strong adsorption of Li atoms on the anodes.

To visualize the effects of adatom adsorption on the charge distribution, we calculate the differential charge density. Fig. 4 clearly shows the charge transfer from the Li atom to the substrates. Further Bader charge analysis show that each Li transfer 0.89, 0.89, and 0.91 electron to C atoms in $PC_2$, $PC_5$, and $PC_6$, respectively, confirming that the adsorption can be mainly attributed to the interaction between Li and C atom.

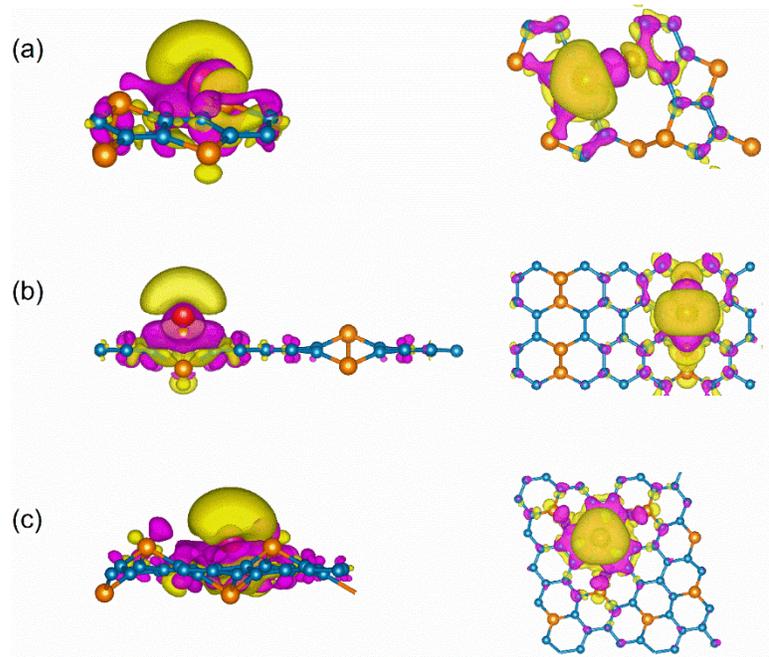

**Fig.4** The differential charge density distribution (Left: side view, and Right: top view) of Li adsorption on (a) $PC_2$, (b) $PC_5$, and (c) $PC_6$. Pink and yellow colors indicate electron accumulation and depletion, respectively.

The charge−discharge rate, which depends on the mobility of the intercalating ions, is another significant character for assessing the capability of an electrode material for rechargeable batteries. Therefore, we simulate the motion of Li atom on the surface of $PC_x$ and investigate the diffusion barriers using the CI-NEB method. Based on the adsorption energy of Li atom on several high-symmetry sites (Fig. S3) in each system, color-filled contour plots are presented in Fig. 5 and S4. From the calculated contour plot of total energy (including adsorbed single Li atom) as a function of adsorption site position, the energetically feasible diffusion paths between the nearest-neighboring lowest-energy adsorption sites ($A_1$ and $A_2$) are selected. For $PC_5$, two possible diffusion pathways are shown in Fig. 5. For Path 1 ($A_1 \rightarrow B \rightarrow A_2$), the diffusion energy barrier is 0.47 eV with diffusion distance of 7.95 Å. For Path 2 ($A_1 \rightarrow C \rightarrow A_2$), the barrier height is 0.57 eV. For $PC_2$ and $PC_6$, the diffusion energy barrier is 0.18 eV and 0.44 eV with diffusion distances of 4.54 Å and 8.98 Å, respectively (Fig. S4). We note that the diffusion barriers on $PC_x$ are comparable to that on typical 2D materials such as graphene (0.33 eV),[51, 52] 1T-MoS$_2$ (0.28 eV),[53, 54] and MoN$_2$ (0.78 eV).[42] The low barriers ensure that Li ions can migrate easily on the surface of our studied systems. Therefore, 2D $PC_x$ systems can be promising electrode materials with fast charge−discharge rate and good rate capability.

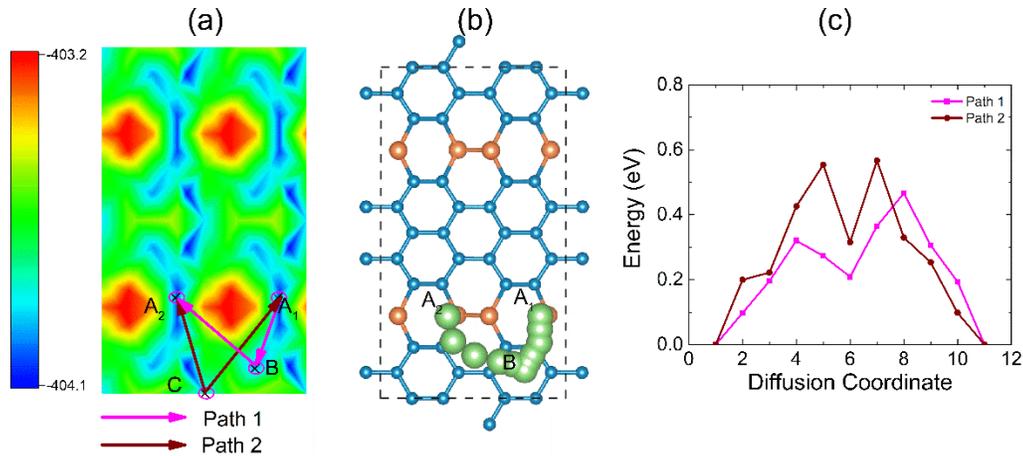

**Fig. 5** (a) Schematic illustration of two possible migration paths of Li diffusion on the PC$_5$ monolayer based on the color-filled contour plots of adsorption energy of Li on the surface of PC$_5$ monolayer, (b) diffusion of single adatoms over PC$_5$ monolayer through the favorable pathway (Path 1), and (c) the corresponding diffusion barrier profiles of Li−PC$_5$.

For practical applications, the storage capacity of batteries is the key indicator for the performance of the electrode materials and the current focus for improvement. To explore the maximum storage capacity for Li atoms, we consider the adsorption with an increasing number of adsorbed Li atoms on both sides of the $PC_x$ monolayers in the 2×2 supercell. To obtain more accurate results, both the atomic positions and lattice constants are fully relaxed for the configurations after the intercalation. The Li atoms are inserted into $PC_x$ systems gradually and randomly until the monolayer reached to its final capacity point. For $PC_2$ monolayer, two Li atoms can be adsorbed in a single cell at most. Otherwise, the structure will collapse (Fig. S5). Thus, the structural deformation may be caused by the intercalation and deintercalation of Li ions in anodes. Therefore, limiting the deformation which is induced by the Li atoms adsorption is necessary. Here, we conduct a series of AIMD simulations to monitor the structural stability of the Li-intercalated $PC_5$ and $PC_6$ nanosheets. Structures with different numbers of adsorbed Li atoms are tested by the simulations to identify the maximum number which does not lead to noticeable deformation of $PC_x$ structures at ambient condition. The AIMD results shown in Fig. 6 present the structures with maximum Li atoms for $PC_5$ and $PC_6$ and all the original bonds are kept completely intact. To be specific, the maximum adatom number is 4.25 per $PC_5$, i.e. $P_8C_{40}Li_{34}$, over which the structure will collapse (Fig. S6). For, $PC_6$, on the other hand, the maximum adatom number is 4.75 (corresponding to chemical formula $P_8C_{48}Li_{38}$), which is judged by the large distance between new adatom and substrate, so that the chemical interactions are too weak to be considered (Fig. S7)[19, 55]. Thus, the intercalation of the Li atoms leads to outstanding capacity of 1251.7 and 1235.9 mAh g$^{-1}$ for PC$_5$ and PC$_6$, respectively. The remarkable storage capacities of the phosphorus carbides can be explained by the quasi-planar structures with abundant adsorption sites and relatively small atomic mass of P and C. The calculated OCV as a function of the adatom content y on the $PC_5$ and $PC_6$ supercell is plotted in Fig. S8. We can see that the curves for PC$_5$ and PC$_6$ show similar trend where the OCV decreases as the number of Li atom y increases at low Li coverage, then converges to a saturated value at higher Li coverage.

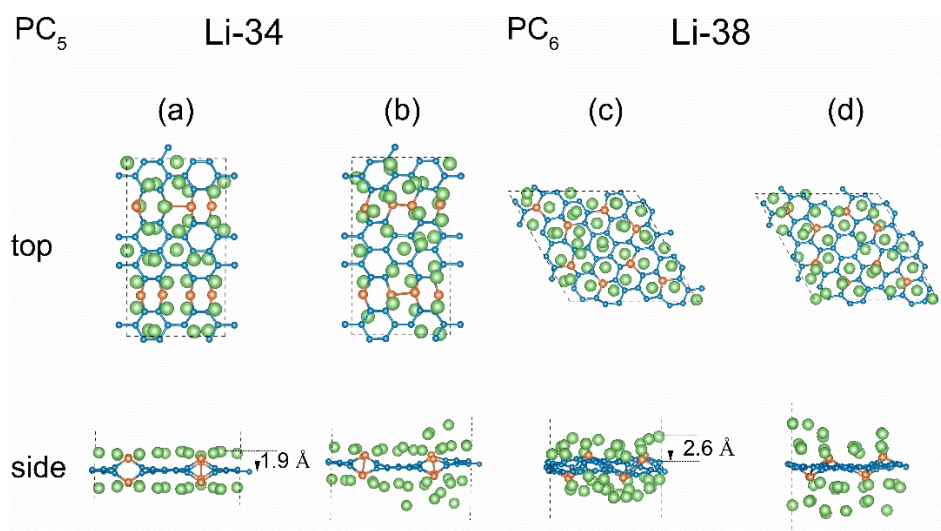

**Fig. 6** Optimized structures before and after AIMD simulations (2 ps, 300 K) of Li-intercalated PC$_5$ and PC$_6$ (2×2 supercell) with maximum numbers of adsorbed Li atoms without deformation, including top and side view of P$_8$C$_{40}$Li$_{34}$ (a) before and (b) after AIMD simulation, and top and side view of P$_8$C$_{48}$Li$_{38}$ (c) before and (d) after AIMD simulation. The Li, P, and C atoms are distinguished by green, orange, and blue color, respectively.

## 4. Conclusion

In summary, we explore the potential applications of 2D $PC_x$ (x=2, 5, 6) monolayers as the anode materials for Li-ion batteries by means of DFT calculations and *ab initio* molecular dynamics simulations. The most energetically favorable diffusion pathways for Li atom in $PC_2$, $PC_5$ and $PC_6$ are identified with considerably low diffusion barriers of 0.18 eV, 0.47 eV and 0.44 eV, respectively, which in turn allows fast charge−discharge rates when they are used as anodes. In addition, $PC_5$ and $PC_6$ monolayer also exhibit high theoretical capacity values of 1251.7 mAh g$^{-1}$ and 1235.9 mAh g$^{-1}$, respectively, which are nearly three times that of graphite. These excellent properties suggest that the dynamically stable $PC_5$ and $PC_6$ monolayers could be promising anode materials. Our results give insightful prospects for further experimental work to explore carbon rich $PC_x$ monolayers as promising electrode material for Li ion batteries.

AUTHOR INFORMATION

**Notes**

The authors declare no competing financial interests.

ACKNOWLEDGMENT

This work was supported by the Research Grants Council of the Hong Kong Special Administrative Region, China (Project No. PolyU152208/18E), the Hong Kong

Polytechnic University (Project Nos. RHA3), and Department of Science and Technology of Guangdong Province (Project No. 2019A050510012). X. L. thanks support from NSFC (No. 11804286) and the fundamental Research Funds for the central Universities.

# Supporting Information

# Theoretical investigation of two-dimensional phosphorus carbides as promising anode materials for lithium-ion batteries

*Ke Fan [a], Yiran Ying [a], Xin Luo [b], and Haitao Huang\* [a]*

[a]Department of Applied Physics, The Hong Kong Polytechnic University, Hung Hom, Kowloon, Hong Kong, P.R. China

[b]School of Physics, Sun Yat-sen University, Guangzhou, Guangdong Province, P.R. China, 510275

AUTHOR INFORMATION

**Corresponding Author**

*E-mail: aphhuang@polyu.edu.hk (H.H.); luox77@mail.sysu.edu.cn (X.L.)

**Table S1.** Structural information (lattice constants, bong length, and bond angle) of $PC_x$ systems

|        | $PC_2$     | $PC_5$           | $PC_6$           |
|--------|------------|------------------|------------------|
| $a$(Å) | 3.96       | 4.39             | 6.69             |
| $b$(Å) | 5.27       | 7.86             | 6.69             |
| Rc-c(Å)| 1.40, 1.43 | 1.39, 1.43, 1.46 | 1.37, 1.42, 1.46 |
| Rc-p(Å)| 1.82       | 1.82             | 1.81             |
| Rp-p(Å)| 2.27       | 2.12             |                  |
| α      | 86.89      | 88.87            | 90.00            |
| β      | 91.70      | 94.31            | 90.00            |
| γ      | 67.95      | 89.99            | 120.00           |

PC$_2$

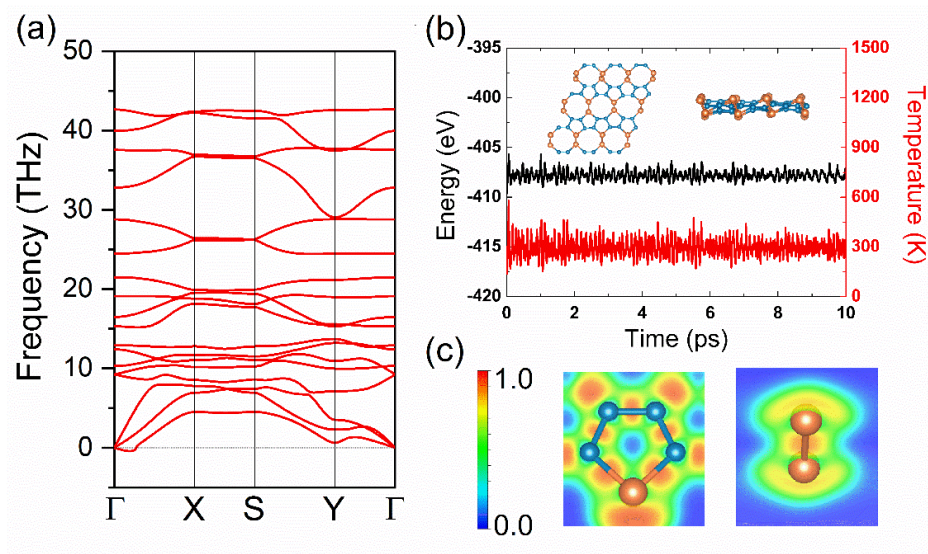

PC$_6$

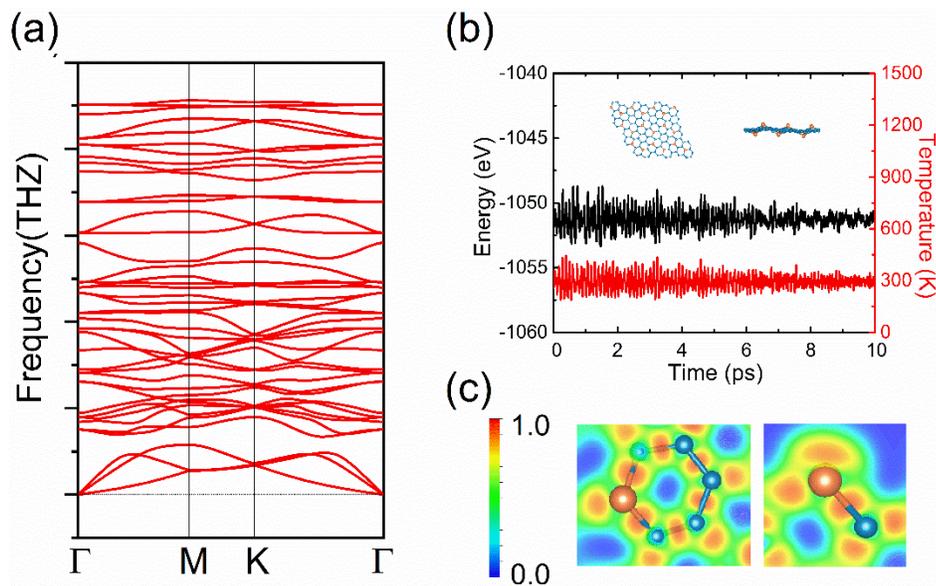

**Fig. S1** Phonon dispersion spectra, AIMD simulation and ELF maps of PC$_2$ and PC$_6$. The P atoms and C atoms are distinguished by orange and blue color, respectively.

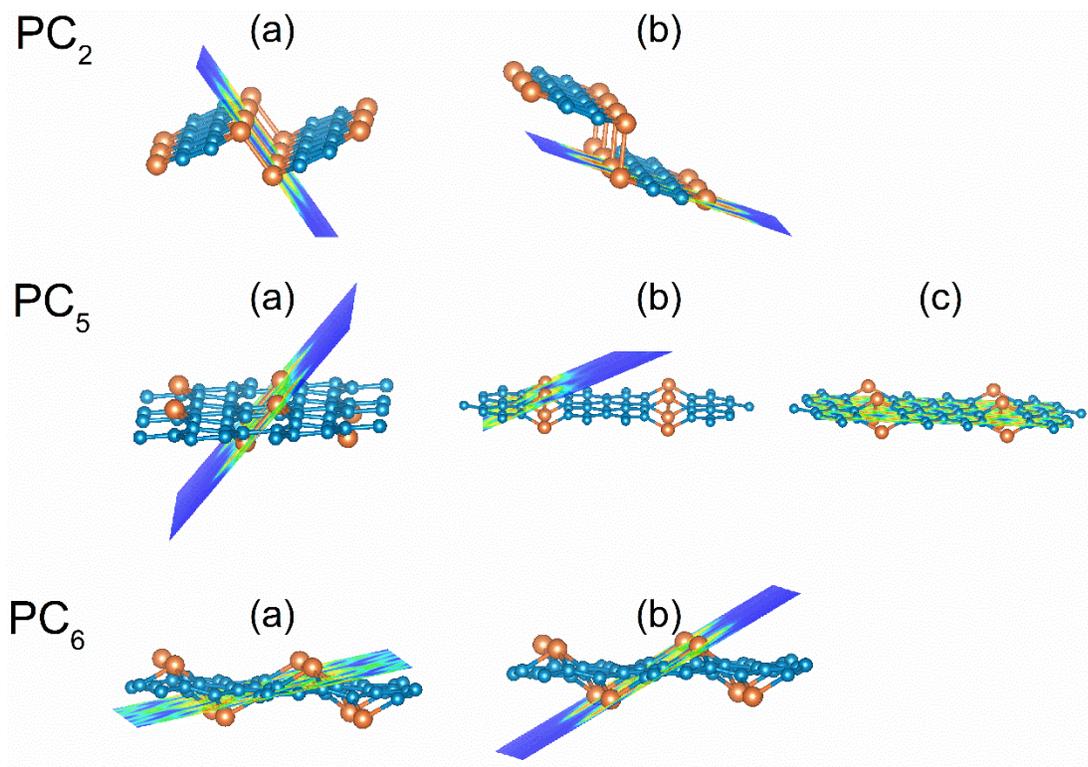

**Fig. S2** Schematic illustration of the crystallographic orientation direction along planes containing specified bonding atoms of $PC_x$ system. The P atoms and C atoms are distinguished by orange and blue color, respectively.

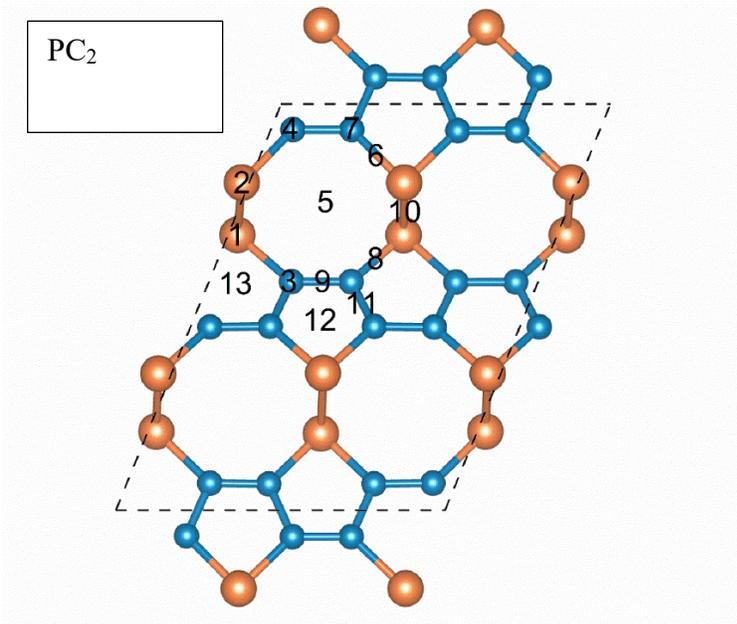

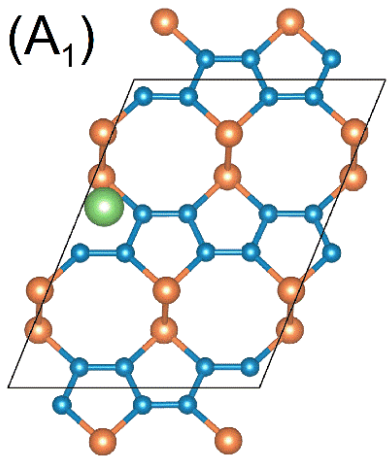 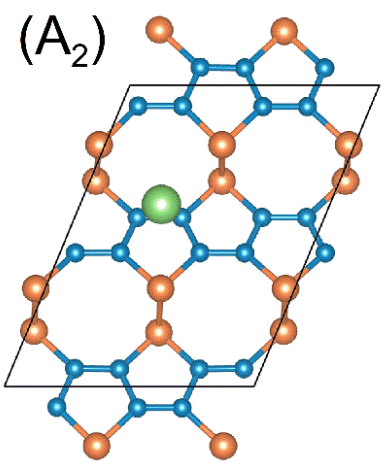 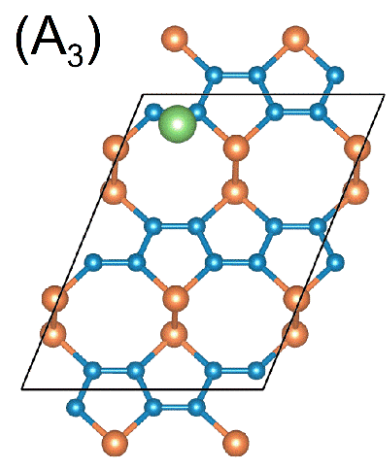

$\Delta E_1 = -0.93 eV$   $\Delta E_2 = -0.87 eV$   $\Delta E_3 = -0.43 eV$

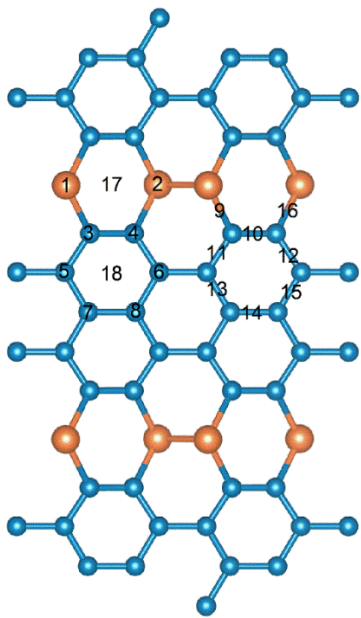

PC$_5$

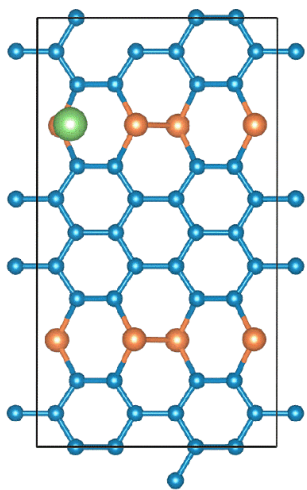
(A$_1$)

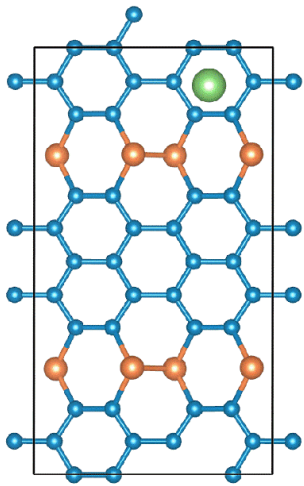
(A$_2$)

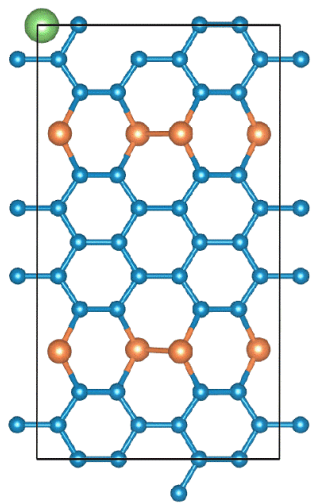
(A$_3$)

$\Delta E_1 = -0.83 eV$       $\Delta E_2 = -0.67 eV$       $\Delta E_3 = -0.57 eV$

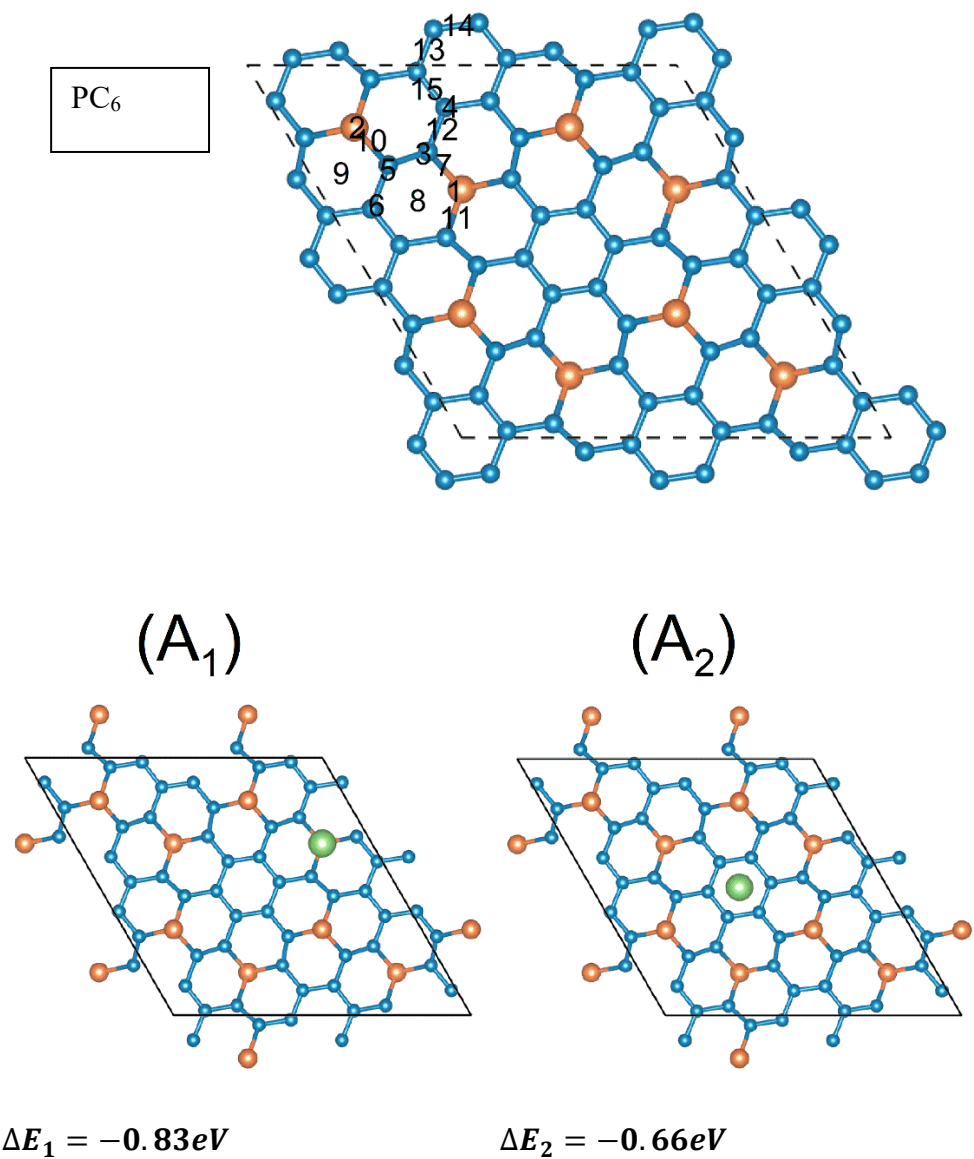

**Fig. S3** Geometric structure of possible adsorption sites on $PC_2$, $PC_5$, and $PC_6$ monolayer. The adsorption energy values for Li atom adsorbed on each site are listed below the corresponding figures. The Li, P, and C atoms are distinguished by green, orange, and blue color, respectively.

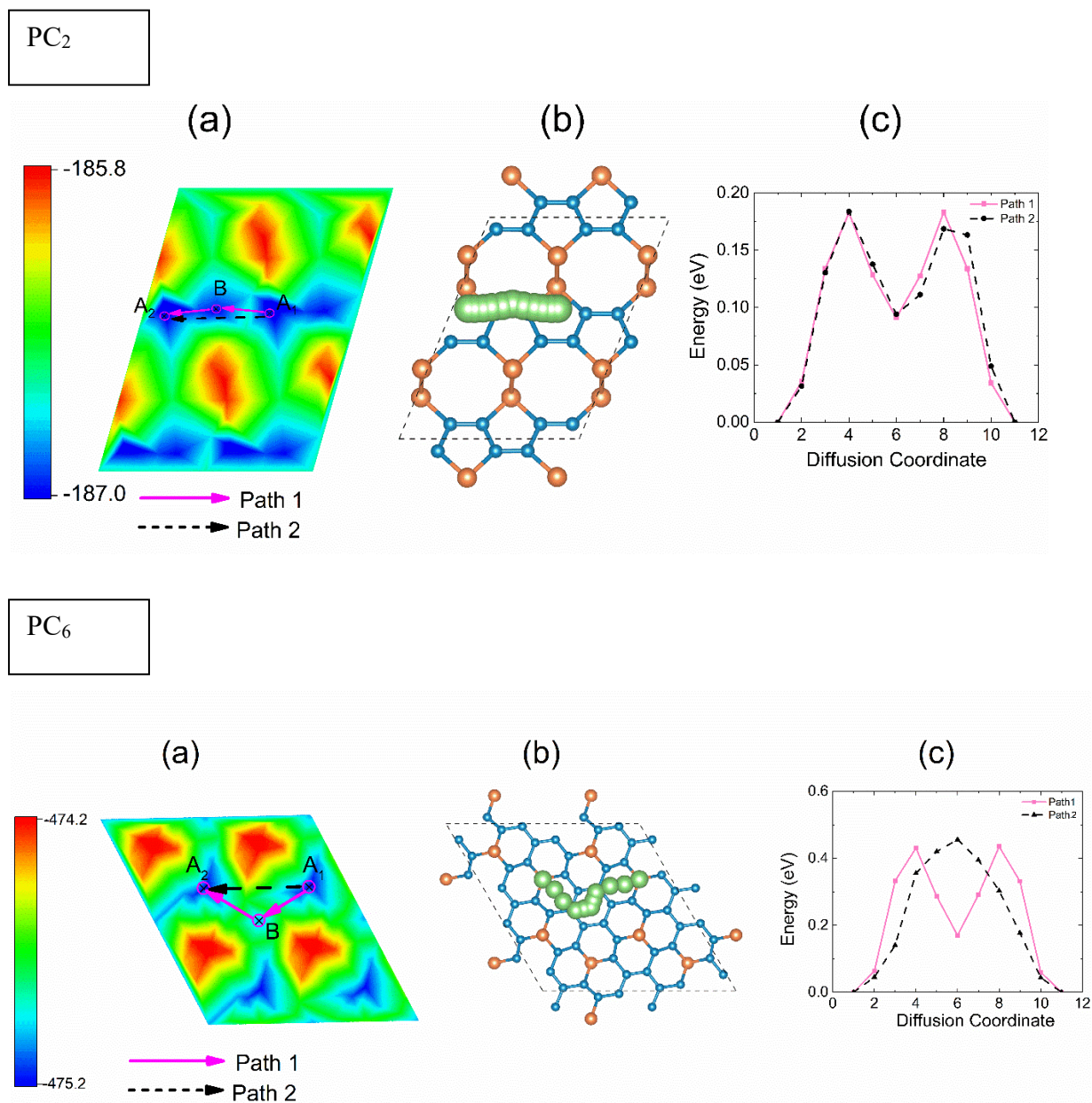

**Fig. S4** (a) Schematic illustration of three possible paths for Li diffusion on the $PC_2$ and $PC_6$ monolayer based on the color-filled contour plots of adsorption energy of single Li atom on the surface; (b) Diffusion of single adatoms over $PC_2$ and $PC_6$ monolayer through the favorable pathways (path 1). (c) The corresponding diffusion barrier profiles of Li−$PC_5$ and Li-$PC_6$ monolayer. The Li, P, and C atoms are distinguished by green, orange, and blue color, respectively.

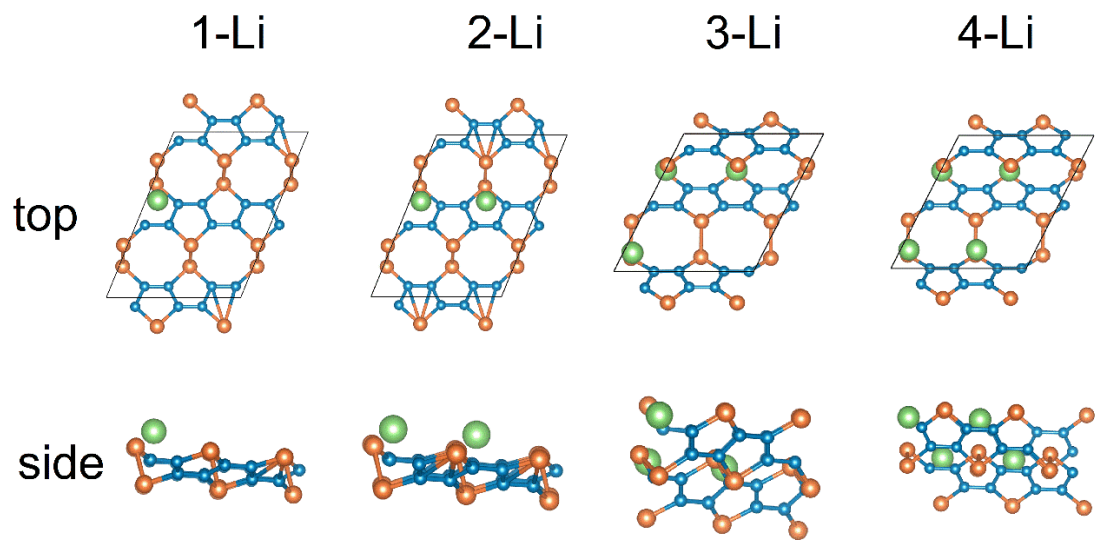

**Fig. S5** Optimized structure of different numbers of Li-intercalated PC$_2$ monolayers (P$_8$C$_{16}$Li, P$_8$C$_{16}$Li$_2$, P$_8$C$_{16}$Li$_3$ and P$_8$C$_{16}$Li$_4$) from top and side view. The Li, P, and C atoms are distinguished by green, orange, and blue color, respectively.

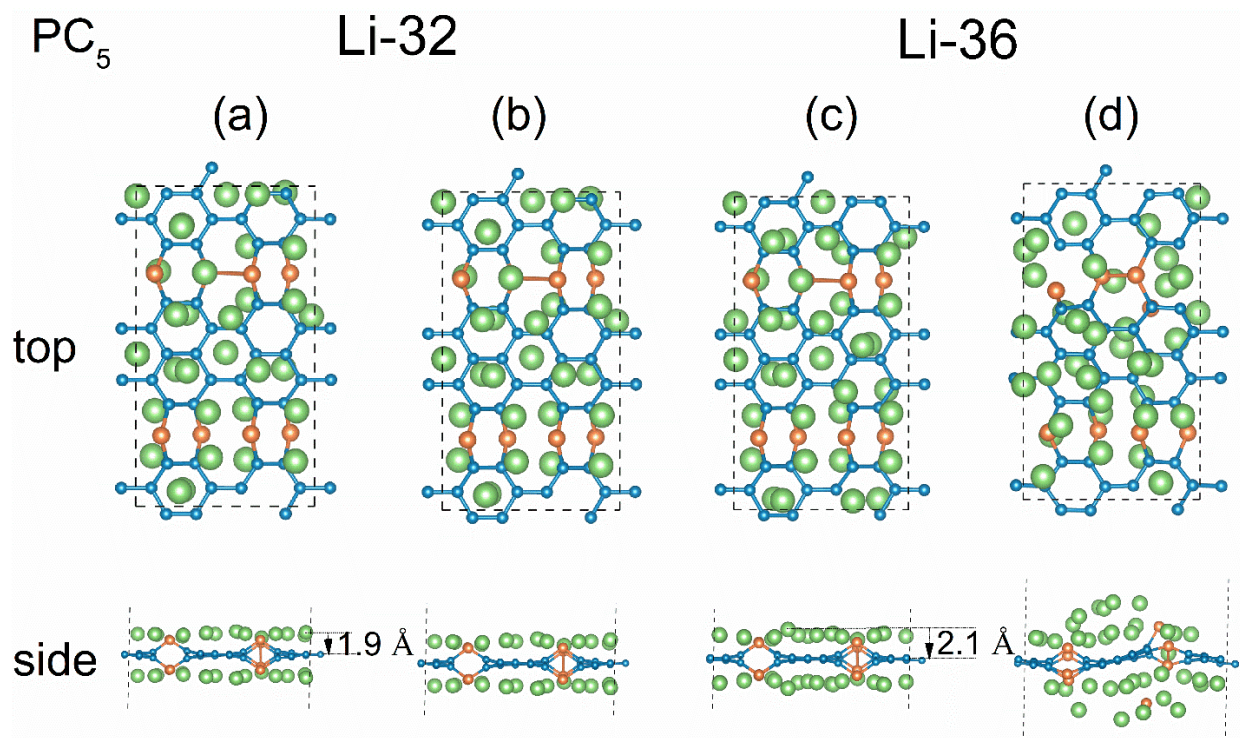

**Fig. S6** Optimized structures of different Li-intercalated $PC_5$ monolayer (2×2 supercell), including $P_8C_{40}Li_{32}$ (a) before and (b) after AIMD simulations, and $P_8C_{40}Li_{36}$ (c) before and (d) after AIMD simulations (300 K). The Li, P, and C atoms are distinguished by green, orange, and blue color, respectively.

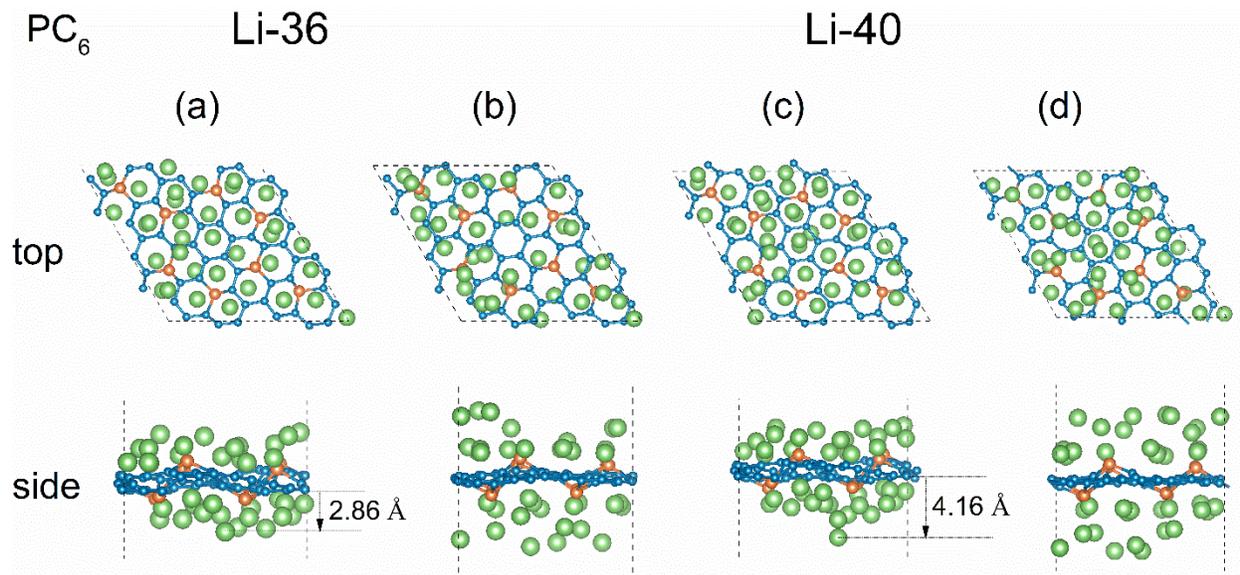

**Fig. S7** Optimized structures of different Li-intercalated PC$_6$ monolayers (2×2 supercell), including P$_8$C$_{48}$Li$_{36}$ (a) before and (b) after AIMD simulations, and P$_8$C$_{48}$Li$_{40}$ (c) before and (d) after AIMD simulations (300 K). The Li atoms, P atoms and C atoms are distinguished by green, orange, and blue color, respectively.

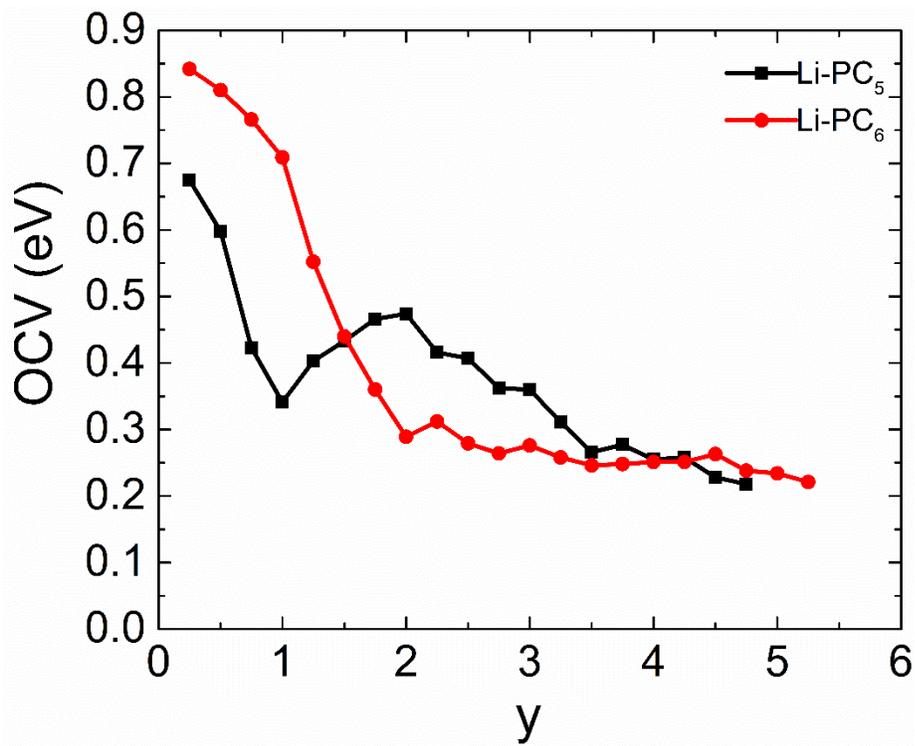

**Fig. S8** Open-circuit voltage (OCV) as a function of adatom content y for PC$_5$ and PC$_6$.